\documentclass[aps,pra,twocolumn,showpacs]{revtex4}

\usepackage{graphicx}
\usepackage{bm}
\usepackage[usenames]{color}

\begin{document}
\title{Fluctuations in the formation time of ultracold dimers from fermionic
atoms}
\author{H.~Uys,  T.~Miyakawa, D.~Meiser, and P.~Meystre}
\affiliation{Optical Sciences Center and Department of Physics\\
The University of Arizona, Tucson, AZ 85721}

\date{\today}

\begin{abstract}
We investigate the temporal fluctuations characteristic of the
formation of molecular dimers from ultracold fermionic atoms via
Raman photoassociation. The quantum fluctuations inherent to the
initial atomic state result in large fluctuations in the passage
time from atoms to molecules. Assuming degeneracy of kinetic
energies of atoms in the strong coupling limit we find that a
heuristic classical stochastic model yields qualitative agreement
with the full quantum treatment in the initial stages of the
dynamics. We also show that in contrast to the association of
atoms into dimers, the reverse process of dissociation from a
condensate of bosonic dimers exhibits little passage time
fluctuations.  Finally we explore effects due to the
non-degeneracy of atomic kinetic energies.
\end{abstract}

\pacs{03.75.Lm, 03.75.Ss, 42.50.Ar}
\maketitle

\section{Introduction}

The coherent formation of ultracold diatomic molecules from
quantum-degenerate bosonic or fermionic atomic gases, via either
Feshbach resonances~\cite{Inouye98} or two-photon Raman
photoassociation~\cite{wynar00}, has witnessed spectacular
developments in recent years and has lead to the first realization
of molecular condensates~\cite{MBEC}. Because in these experiments
the molecular field is initially in a vacuum state, it is to be
expected that quantum fluctuations play a dominant role in the
early stages of molecule formation. These fluctuations manifest
themselves in the quantum statistics of the resulting molecular
field, and also in the time that it takes for the number of
generated molecules to reach a specific value, the so-called
passage time statistics. This problem is closely related to other
situations where quantum (or thermal) fluctuations trigger a
system to undergo a transition away from a dynamically unstable
state. One such example familiar from quantum optics is
superradiance \cite{dicke54} (or strictly speaking
superfluorescence), a situation where an ensemble of two-level
systems initially in their excited electronic state and coupled to
the electromagnetic field vacuum undergoes a transition
characterized by the emission of an intense light pulse. One
important difference is that in superfluorescence experiments the
atoms are normally coupled to a continuum of modes of the
radiation field, practically leading to an irreversible decay to
their ground state, while in the problem at hand the molecular
field is to a good approximation single-mode, leading to
reversible dynamics.

One important way to characterize the dynamics of molecule
formation is by way of the so-called passage time, which is
defined as the time it takes to produce a predetermined number of
molecules. The goal of this paper is to study the passage time
statistics resulting from the initial quantum fluctuations of the
atomic matter-wave field. We also compare this situation with the
dynamics of dissociation of a molecular condensate into fermionic
atomic pairs, showing significant qualitative differences between
the two cases.

The paper is organized as follows: Section II establishes our
notation, presents our model, and shows that the application of an
Anderson mapping \cite{anderson} leads to the description of
photoassociation of fermions in terms of the inhomogeneously
broadened Tavis-Cummings model of quantum optics. Section III
concentrates on a ``homogeneously broadened'' version of this
model that neglects the spread in fermion energies, an
approximation  shown to be valid for sufficiently small numbers of
atoms. There we also discuss an approximate stochastic classical
description \cite{haake79, haake81a} that yields a satisfactory
qualitative agreement with the full quantum results for short
enough times. This section concludes by comparing the passage time
statistics associated with photoassociation and the reverse
process of photodissociation. The results of a numerical analysis
of the full, inhomogeneously broadened model are presented in
section IV. Finally section V is a summary and outlook.

\section{The model}

In typical experiments that produce molecules via Feshbach
resonance the magnetic field is swept across the
resonance. In the strong coupling regime $k_F a \ge 1$ in the
vicinity of the resonance, where $k_F$ is the Fermi wave number
and $a$ is the s-wave scattering length, the interpretation of the
molecular state is subject to conceptual difficulties stemming
from the dressing of the "bare" molecular state by atom pairs in
the open channel ~\cite{bruun04}. Furthermore, since the binding
energy of the molecules is very small, of the order $10^{-11}$eV,
they are larger than the interatomic separation. Another
consequence of the almost vanishing binding energy is that nearly
every time-dependent process becomes nonadiabatic with respect to
the time scale set by the inverse of the binding energy. To avoid
these difficulties we restrict our considerations in this paper to
the description of a quantum-degenerate gas of fermionic atoms of
mass $m_f$ and spin $\alpha=\uparrow, \downarrow$, coupled
coherently to bosonic molecules of mass $m_b = 2m_f$ and zero
momentum via \textit{photoassociation} rather than Feshbach
resonance.

Neglecting collisions between fermions and assuming that for short
enough times the molecules can be described by a single-mode
bosonic field, this system can be described by the boson-fermion
model Hamiltonian

\begin{eqnarray}
\label{Hamil1} H &=& \sum_{k}
\frac{1}{2}\hbar\omega_k\left(\hat{c}^\dagger_{k\uparrow}\hat{c}_{k\uparrow}
+
\hat{c}^\dagger_{-k\downarrow}\hat{c}_{-k\downarrow}\right)+\hbar\omega_b\hat{b}^\dagger
\hat{b}\nonumber\\
&+& \hbar\chi\sum_k\left(\hat{b}^\dagger
\hat{c}_{k\uparrow}\hat{c}_{-k\downarrow} +
\hat{b}\hat{c}^\dagger_{-k\uparrow}\hat{c}^\dagger_{k\downarrow}\right),
\end{eqnarray}
where $\hat{b}^\dagger, \hat{b}$ are molecular bosonic creation
and annihilation operators and $\hat{c}_{k\alpha}^\dagger,
\hat{c}_{k\alpha}$ are fermionic creation and annihilation
operators describing atoms of momentum $\hbar k$ and spin
$\alpha$. The first and second terms in Eq. (\ref{Hamil1})
describe the energy, $\omega_k=\hbar^2k^2/m_f$, of the atoms,  and
the detuning energy of the molecules respectively, and the third
term describes the photoassociation of pairs of atoms of opposite
momentum into molecules.

Introducing the pseudo-spin operators \cite{anderson}
\begin{eqnarray}
    \hat{\sigma}^z_k &=&
\frac{1}{2}(\hat{c}^\dagger_{k\uparrow}\hat{c}_{k\uparrow} +
    \hat{c}^\dagger_{-k\downarrow}\hat{c}_{-k\downarrow}-1),\nonumber \\
    \hat{\sigma}^+_k &=&(\hat{\sigma}^-_k)^\dagger=
    \hat{c}^\dagger_{-k\downarrow}\hat{c}^\dagger_{k\uparrow},
\label{A-mapping}
\end{eqnarray}
which are easily seen to obey the SU(2) commutation relations
\begin{eqnarray}
\left[\hat{\sigma}^+_{k},\hat{\sigma}^-_{k'}\right]
&=&2\delta_{kk'}\hat{\sigma}^z_k\\
\left[\hat{\sigma}^z_{k},\hat{\sigma}^\pm_{k'}\right]
&=&\pm\delta_{kk'}\hat{\sigma}^\pm_k,
\end{eqnarray}
where $\delta_{kk'}$ is the Kronecker delta function, the
Hamiltonian (\ref{Hamil1}) becomes, within an unimportant constant
\cite{barankov04,Dominic04},
\begin{equation}
\label{Hamil2} H = \sum_k \hbar\omega_k\hat{\sigma}^z_k
+\hbar\omega_b\hat{b}^\dagger \hat{b} +
\hbar\chi\sum_k\left(\hat{b}^\dagger\hat{\sigma}^-_k +
\hat{b}\hat{\sigma}^+_k\right).
\end{equation}
This Hamiltonian is known in quantum optics as the inhomogeneously
broadened (or non-degenerate) Tavis-Cummings model~\cite{tavis68}.
It describes the coupling of an ensemble of two-level atoms to a
single-mode electromagnetic field. Hence the mapping
(\ref{A-mapping}) establishes the formal analogy between the
problem at hand and Dicke superradiance, with the caveat already
mentioned that we are dealing with a single bosonic
mode~\cite{barankov04,Dominic04,javanainen04,andreev04,vardi04,Taka05}.
Instead of real two-level atoms, pairs of fermionic atoms are now
described as effective two-level systems whose ground state
corresponds to the absence of a pair, $|g_k\rangle
=|0_{k\uparrow},0_{-k\downarrow}\rangle$ and the excited state to
a pair of atoms of opposite momenta, $|e_k\rangle =
|1_{k\uparrow},1_{-k\downarrow}\rangle$.

The initial condition of the superradiance problem is a sample of
inverted two-level atoms. It corresponds in the present case to
the initial atomic state
    \begin{equation}
    |F\rangle=\prod_{k} \hat{\sigma}^+_k |0\rangle,
    \end{equation}
where the product is taken up to the Fermi surface for $T=0$,
while the molecular field is in the vacuum state $|0\rangle$. We
concentrate in the following on times short enough that the atomic
sample remains essentially undepleted and it is sufficient to
consider only fermionic levels up to the Fermi surface in
Eq.~(\ref{Hamil2}).

\section{Degenerate model}

We consider first the simplified situation of a degenerate model
in which the inhomogeneous broadening due to the spread in atomic
kinetic energies is ignored. This is justified provided that
these energies are small compared to the atom-molecule coupling
energy, $\beta=\epsilon_F/(\hbar \chi)\ll1$, where $\epsilon_F$ is
the Fermi energy.  This approximation is
the analog of the homogeneous broadening limit of quantum optics,
and of the Raman-Nath approximation in atomic diffraction.  As we
will find it is valid only for relatively
small ($\sim 10^2-10^3$) particle
numbers, but the model exhibits the essential physics.

The atom-molecule coupling has been estimated
~\cite{heinzen00,Foot1} for the case of $^{87}$Rb to be
$\chi\sqrt{V} \approx 7.6\times 10^{-7}$ m$^{3/2}$s$^{-1}$, so
that $\beta=\epsilon_F/\hbar\chi
\approx 10^{-2}N^{7/12}.$ In the last term we have related the
Fermi energy and volume to the oscillator frequency $\omega_{\rm
ho}$ of a spherically symmetric harmonic trap via
$\epsilon_F\approx N^{1/3}\hbar\omega_{ho}$ and $V \approx N^{1/2}
a_{\rm osc}^3 $ where $a_{\rm osc}=\sqrt{\hbar/m\omega_{ho}}$ is
the oscillator length. Typical experiments use traps with
$\omega_{\rm ho} \approx 100$ Hz.  In the case of $^{87}$Rb the
trap should contain at most $\sim 10^2 - 10^3$ atom pairs to be in
this regime, i.e. to have $\beta \lesssim 1$. For larger samples,
it is necessary to account for the inhomogeneous broadening of the
sample, a situation that we consider in the next section. For any
atom numbers, the characteristic time scale is given
$\tau_p=1/\chi \sqrt{N}\simeq 5.8 \times 10^{-3} N^{-1/4}$s
for parameters used in this paper.

\subsection{Quantum description}

Limiting for now our considerations to small atomic samples, we
approximate all $\omega_k$'s by $\omega_F$ and introduce the
collective pseudo-spin operators
    \begin{eqnarray}
    \hat{S}_z = \sum_k \hat{\sigma}^z_k, \nonumber\\
    \hat{S}^\pm = \sum_k \hat{\sigma}^\pm_k,
    \end{eqnarray}
which again obey SU(2) commutation relations, yielding the standard
Tavis-Cummings Hamiltonian~\cite{tavis68,Taka05}
\begin{equation}
\label{Hamil3} H = \hbar\omega_F \hat{S}_z +\hbar\omega_b
\hat{b}^{\dagger}\hat{b} +\hbar\chi(\hat{b}\hat{S}^{+} +
\hat{b}^{\dagger}\hat{S}^-).
\end{equation}
This Hamiltonian conserves the total spin operator $\hat{\bm S}^2$,
which, by using the pseudo-spin commutation relations, can be
expressed as
    \begin{equation}
    \label{S}
    \hat{\bm S}^2 = \hat{S}^+\hat{S}^- + \hat{S}_z(\hat{S}_z-1),
    \end{equation}
with
    \begin{equation}
    \hat{\bm S}^2|F\rangle= S(S+1)|F\rangle    =\frac{N}{2}\left(\frac{N}{2}+1\right)|F\rangle,
    \end{equation}
so that $S=N/2$. Here
    \begin{equation}
    N =\hat{b}^{\dagger}\hat{b} +
\sum_{k}(\hat{c}^\dagger_{k\uparrow}\hat{c}_{k\uparrow}
    + \hat{c}^\dagger_{-k\downarrow}\hat{c}_{-k\downarrow})/2  =
    {\hat n}_b + {\hat n}_p
    \end{equation}
is the total number of molecules and atomic pairs, which is
conserved by the Hamiltonian (\ref{Hamil1}). From the definition
of $S_z$ we also have that
\begin{equation}
\label{sznb} \hat{S}_z = \frac{1}{2}\left (2{\hat n}_p -N\right ) =
\frac{N}{2}-\hat{n}_b =
\frac{1}{2} \left({\hat n}_p - {\hat n}_b\right ),
\end{equation}
hence $\hat{S}_z$ measures the difference in the numbers of atom pairs
and molecules.

Introducing for convenience the joint coherence operators
    \begin{eqnarray}
    \hat{J}_x &=& (\hat{b}\hat{S}^+ + \hat{b}^\dagger \hat{S}^-)/2, \nonumber \\
    \hat{J}_y &=& (\hat{b}\hat{S}^+ - \hat{b}^\dagger \hat{S}^-)/2i,
    \end{eqnarray}
yields the Heisenberg equations of motion
\begin{eqnarray}
\label{heisen1}
    \dot{\hat{n}}_b &=&-2\chi \hat{J}_y, \\
\label{heisen2}    \dot{\hat{J}_x} &=& \delta \hat{J}_y \\
\label{heisen3}    \dot{\hat{J}_y} &=& -\delta \hat{J}_x-\chi\left (2
\hat{S}_z\hat{n}_b +
\hat{S}^+\hat{S}^-\right ),
\end{eqnarray}
where $\delta = \omega_b-\omega_F$, so that $2\chi \hat{J}_x + \delta
\hat{n}_b$ is a constant of motion.

\begin{figure}
\begin{center}
\includegraphics[width=8cm,height=4.5cm]{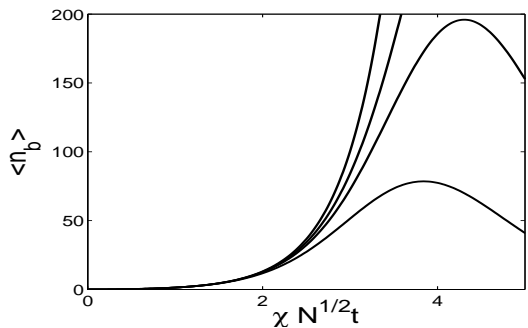}
\end{center}
\caption{Short-time dynamics of $\langle {\hat n}_b\rangle$. From
left to right, the curves give the linearized solution
(\ref{nbt2}) and the full quantum results for $N=500$, $N=250$,
and $N=100$, respectively. } \label{fig1}
\end{figure}

\begin{figure}
\begin{center}
\includegraphics[width=8cm,height=4.5cm]{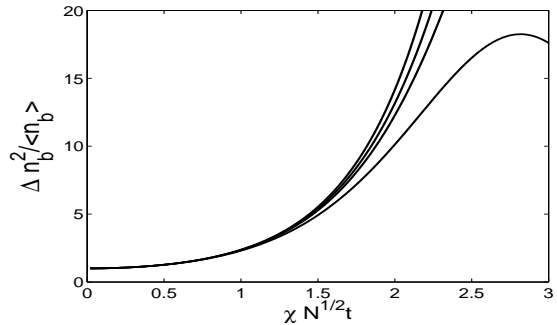}
\end{center}
\caption{Short time dynamics for $\Delta n_b^2/\langle {\hat
n}_b\rangle$. From left to right, the four curves give the
linearized solution (\ref{dnbt2}) and the full quantum results for
$N=500$, $N=250$, and $N=100$ respectively. } \label{fig2}
\end{figure}

In the following, we confine our discussion to the case of
$\delta = 0$ for simplicity.
We thus neglect the contribution of $J_x$ in Eq. (\ref{heisen3}).
In order to obtain
an analytical solution valid for short times for
$\langle\hat{n}_b\rangle$, where $\langle\,\rangle$ indicates the
expectation value, and assuming the initial state $|F\rangle$, we
keep only terms of order $\hat{n}_b$ on the right in Eq.
(\ref{heisen3}). Using Eqs.~(\ref{S}) and (\ref{sznb}), we
reexpress $\hat{S}^{+}\hat{S}^{-}$ as
\begin{equation}
\label{spm}
\hat{S}^+\hat{S}^- = -\hat{n}_b^2 + (2S-1)\hat{n}_b + \hat{\zeta}^+\hat{\zeta}^-,
\end{equation}
where we have introduced for convenience the operator
\begin{equation}
\hat{\zeta}^+\hat{\zeta}^- = \hat{{\bm S}}^2 - S(S-1).
\end{equation}
Substituting then Eq. (\ref{spm}) into Eq. (\ref{heisen3}) and
dropping the term proportional to ${\hat n}_b^2$, we have for the
early stages of molecule formation
    \begin{equation}
    \label{jy2} \dot{\hat{J}_y} \approx -2\chi N
    \hat{n}_b - \chi \hat{\zeta}^+\hat{\zeta}^-.
    \end{equation}
Differentiating Eq. (\ref{heisen1}), taking its expectation
value, and substituting Eq. (\ref{jy2}) into the resulting form
yields
\begin{equation}
\ddot{\hat{n}}_b \approx \chi^2\left
(4N\hat{n}_b+2\hat{\zeta}^+\hat{\zeta}^-\right)
\end{equation}
For the initial state $|F\rangle$ this has the solution
\begin{equation}
\label{nbt2} \langle {\hat n}_b(t)\rangle \approx \left
(\langle\hat{\zeta}^+\hat{\zeta}^-\rangle/N \right )\sinh^2{(\chi\sqrt{N}t)}.
\end{equation}
Similarly, the variance of the molecule number distribution is found
to be
\begin{eqnarray}
\label{dnbt2} \nonumber \Delta n_b^2(t)&=& \langle {\hat n}_b^2(t)\rangle -
\langle {\hat n}_b(t)\rangle^2\\
&\approx& \left
(\langle\hat{\zeta}^+\hat{\zeta}^-\rangle/8 N
\right)\left(\cosh{(4\chi\sqrt{N}t)}-1\right).
\end{eqnarray}

\begin{figure}
\begin{center}
\includegraphics[width=8cm,height=4.5cm]{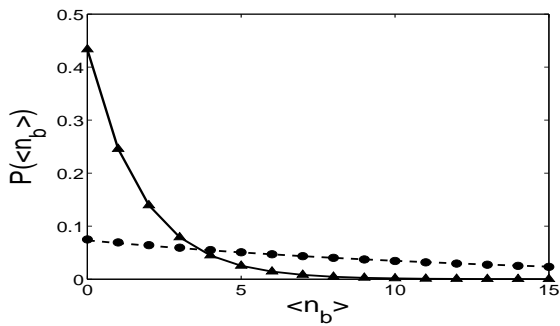}
\end{center}
\caption{Molecule number statistics at $\chi\sqrt{N}t = 1.0$:
solid line - full quantum result, triangles - thermal distribution,
and $\chi\sqrt{N}t = 2.0$: dashed line - full quantum result,
circles - thermal distribution, for $N=500$.} \label{fig3}
\end{figure}

Figure~\ref{fig1} compares the average molecule number $\langle
{\hat n}_b\rangle$ and Fig.~\ref{fig2} the normalized variance $\Delta
n_b^2/\langle \hat{n}_b\rangle$ from Eqs. (\ref{nbt2}) and
(\ref{dnbt2}) respectively, with the full quantum solution
obtained by direct diagonalization of the
Hamiltonian~(\ref{Hamil3}). Both approaches agree within $
5\%$ until about $20\%$ of the population of atom pairs has been
converted into molecules. Note that
\begin{equation}
\lim_{\chi\sqrt{N}t\rightarrow
0}{\Delta n_b^2/\langle \hat{n}_b\rangle}=1.
\end{equation}
This is indicative of the fact that for short times the molecule
field is thermal in character, see for example \cite{Dominic04}
and references therein. This is further confirmed by a comparison
of the molecular number statistics to a thermal distribution, as
illustrated in Fig.~\ref{fig3}.

\subsection{Classical stochastic description}

Eq.(\ref{jy2}) shows that for the initial state $|F\rangle$ the
source of the molecular field is the operator
$\hat{\zeta}^+\hat{\zeta}^-$.  This is the non-vanishing part of
the second-order moment $\langle \hat{S}^+\hat{S}^-\rangle$, which
is a measure of fluctuations of the atomic field. Since the
molecular field is initially in a vacuum, its growth is therefore
triggered solely by these quantum fluctuations.

It is oftentimes possible to simulate the effects of quantum
fluctuations by averaging over a large number of classical
trajectories triggered by random noise. To implement such an
approach, we first observe that for $N\gg 1$ the higher-order
moments of the pseudo-spin operators $\hat{S}^{\pm}$ factorize
approximately into a sum of products of second-order moments
\cite{ reed62,mandel95}, e.g.,
\begin{equation}
\label{momentth} \langle \hat{S}^+\hat{S}^+\hat{S}^-\hat{S}^- \rangle = 2\langle
\hat{S}^+\hat{S}^-\rangle \langle \hat{S}^+\hat{S}^- \rangle + O(S^{-1}).
\end{equation}
We then proceed by replacing the quantum operator $\hat{S}^+$ by a
stochastic classical variable $s^+$ and assuming that its
fluctuations obey random Gaussian statistics, the probability
distribution of $s^+$ being given by
\begin{equation}
\label{distrib} p(s^+) = \frac{1}{\sqrt{2\pi}\Delta \hat{S}^{+}}
\exp\left [-|s^+|^2/2(\Delta \hat{S}^{+})^2\right ]
\end{equation}
where the variance $(\Delta \hat{S}^{+})^2$ is adjusted to its quantum value
    \begin{equation}
    (\Delta \hat{S}^{+})^2 = \langle |\hat{S}^+|^2
    \rangle - \langle \hat{S}^{+} \rangle^2=N
    \end{equation}
for the initial state $|F\rangle$.

The resulting passage time statistics $\rho(\tau)$ can be determined
numerically by first obtaining classical trajectories for $\langle
\hat{n}_b (t)\rangle$ using Eq. (\ref{nbt2}) and assuming that the
initial values of $s^+$ follow the distribution (\ref{distrib}).
Taking into account that
each choice of $|s^+|^2$ within a
differential element $ds^+$ maps $\langle \hat{n}_b(t)\rangle$ in
such a way that it reaches a fixed reference value $n_b^{\rm ref}$
after a uniquely determined passage time $\tau$ within the
differential element $d\tau$, it follows that the probability of a
particular value of $s^+$ is equal to the probability of measuring
$\langle \hat{n}_b\rangle =n_b^{\rm ref}$ at that time $\tau$
    \begin{equation}
    |p(s^+)|ds^+(\tau,n_b^{\rm ref}) = \rho(\tau)d\tau =
    |p(s^+(\tau))f(\tau, n_b^{\rm ref})|d\tau,
    \end{equation}
the differential elements $ds^+$ and $d\tau$ being related by
    \begin{equation}
    f(\tau,n_b^{\rm ref}) = |ds^+(\tau,n_b^{\rm ref})/d\tau|.
    \end{equation}
By fixing $\langle \hat{n}_b(\tau)\rangle=n_b^{\rm ref}$ one can
invert Eq. (\ref{nbt2}) which upon differentiation with respect to
$\tau$ gives
\begin{equation}
f(\tau,n_b^{\rm ref}) = \chi
N\sqrt{n_b^{\rm ref}}\frac{\cosh{(\chi\sqrt{N}\tau)}}{\sinh{^2(\chi\sqrt{N}\tau)}}.
\end{equation}
We thus obtain an analytical expression for the passage time
distribution $\rho(\tau)$ as
\begin{eqnarray}
\label{analytic1} \rho(\tau) &=&
\left ( \chi\sqrt{ \frac{2n_b^{\rm ref}N}{\pi}} \right )
\frac{\cosh{(\chi\sqrt{N}\tau)}}{\sinh{^2(\chi\sqrt{N}\tau)}}\nonumber\\
&\times&\exp{\left(-\frac{n_b^{\rm
ref}}{2\sinh{^2(\chi\sqrt{N}\tau)}}\right )}.
\end{eqnarray}

\begin{figure}
\begin{center}
\includegraphics[width=8cm,height=4.5cm]{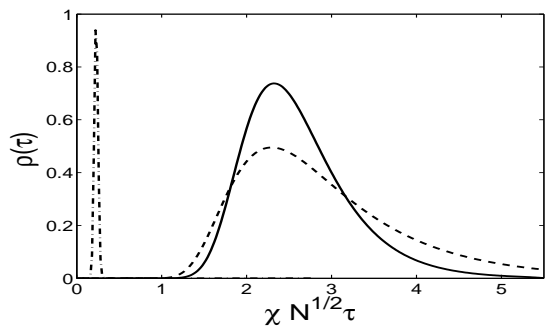}
\end{center}
\caption{Passage time distribution for converting 5$\%$ of the
initial population consisting of only atoms (molecules) into
molecules (atoms) for $N = 500$. For initially all atoms: solid
line - full quantum description; dashed line - classical
stochastic model. For initially all molecules: dot-dashed line -
full quantum result}. \label{fig4}
\end{figure}

The dashed line in Fig.~\ref{fig4}, obtained from
Eq.~(\ref{analytic1}), shows the distribution of passage times
required to produce a normalized molecule number $n_b^{\rm ref}/N
= 0.05$ from a sample initially containing $N=500$ pairs of atomic
fermions. It should be compared to the solid line, which is the
result of the full quantum dynamics. The classical result
reproduces qualitatively the broad and asymmetric distribution of
the quantum solution. However, it does not reproduce well the
leading and trailing edges of the distribution, which depend on
the higher order moments of the classical field $s^+$ and are
poorly treated by the assumption of Gaussian noise. In addition
the continuous distribution of the classical model fails to
properly describe the dynamics of the system at very low molecule
numbers when the discrete nature of molecule number is more
important.

\subsection{Photodissociation}

\begin{figure}
\begin{center}
\includegraphics[width=8cm,height=4.5cm]{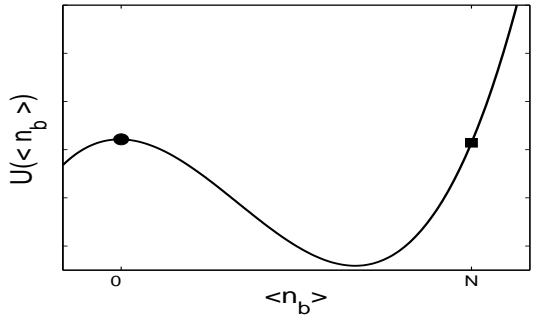}
\end{center}
\caption{Effective potential for a system with $N\gg1$. The
circle (square) corresponds to an initial state with all fermionic
atoms (molecules). The part of the potential for $n_b<0$ is unphysical.}
\label{fig5}
\end{figure}

The passage time distribution for the photoassociation of
fermionic atoms into molecules differs sharply from its
counterpart for the reverse process of photodissociation from a
molecular condensate into fermionic atom pairs, which is plotted
as the dot-dashed line in Fig.~\ref{fig4}. In contrast to photoassociation,
this latter process suffers significantly reduced fluctuations.
One can gain an intuitive understanding of this difference by
treating the short-time molecular population classically,
$\langle{\hat n}_b\rangle \rightarrow n_b$. Within this
approximation, the Heisenberg equations of motion
(\ref{heisen1})-(\ref{heisen2}) can be recast in the form of a
Newton equation~\cite{Taka05}
\begin{equation}
\label{newton} \frac{d^2 n_b}{dt^2} = -\frac{d U(n_b)}{dn_b},
\end{equation}
where the effective potential $U(n_b)$ is given by
    \begin{equation}
    U(n_b)= \frac{1}{4}N^2(N+3) -2Nn_b-(2N-1)n_b^2+2n_b^3
    \end{equation}
and is cubic in $n_b$, see Fig.~\ref{fig5}.

In case the system is initially composed solely of fermionic
atoms, $n_b(0)=0$, the initial state is dynamically unstable, with
fluctuations having a large impact in the build-up of $n_b$. In
contrast, when it consists initially solely of molecules, $n_b=N$,
the initial state is far from the point of unstable equilibrium,
and $n_b$ simply ``rolls down'' the potential in a manner largely
insensitive to quantum fluctuations. This is a consequence of the
fact that the bosonic initial state provides a mean field that is
more amenable to a classical description.

\section{Non-degenerate model}

Although the degenerate model offers the benefit of allowing
analytical solutions and an intuitive physical interpretation in
the short time limit, we have seen that it is only valid for
relatively modest values of $N$. Current experiments, however,
generally trap $\sim 10^5 - 10^6$ atoms, in which case it is
important to properly account for the atomic kinetic energies
within the Fermi sea.

\begin{figure}
\begin{center}
\includegraphics[width=8cm,height=4.5cm]{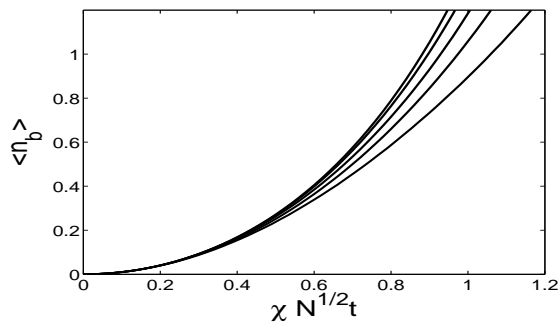}
\end{center}
\caption{Average molecule number as a function of time for various
values of the coupling strength. From left to right: degenerate
analytical, full quantum treatment for $\beta = 0.1$; short-time
treatment for $\beta = 0.1$, short-time treatment for $\beta =
10$, full quantum treatment for $\beta = 10$. In this example we
used $N = 18$.} \label{fig6}
\end{figure}

From the Hamiltonian (\ref{Hamil2}), we readily obtain the
Heisenberg equations of motion:
\begin{eqnarray}
\label{nbk}
\frac{d \hat{n}_b}{dt}
&=&-2\chi \sum_k  \hat{j}_k^y,\\
\label{jkx}
\frac{d \hat{j}_k^x}{dt}
&=& \delta_k  \hat{j}_k^y, \\
\nonumber
\frac{d \hat{j}_k^y}{dt}
&=& -\delta_k  \hat{j}_k^x-2\chi
 \hat{n}_b\hat{\sigma}_k^z - \frac{1}{2}\chi
\sum_{k'} \left
(\hat{\sigma}^+_k\hat{\sigma}^-_{k'}+\hat{\sigma}^+_{k'}\hat{\sigma}^-_{k}\right)\\
&\approx& -\delta_k  \hat{j}_k^x-2\chi \hat{n}_b\hat{\sigma}_k^z -
\chi\hat{\sigma}^+_k\hat{\sigma}^-_{k},\label{jky}
\end{eqnarray}
where we have defined
\begin{eqnarray}
\hat{j}_k^x &=& \frac{\hat{b}\hat{\sigma}^+_k + \hat{b}^\dagger
\hat{\sigma}^-_k}{2},\\
\hat{j}_k^y &=& \frac{\hat{b}\hat{\sigma}^+_k - \hat{b}^\dagger
\hat{\sigma}^-_k}{2i},
\end{eqnarray}
and $\delta_k = \omega_b - \omega^f_k$. In Eq. (\ref{jky}) we have also made the
approximation
\begin{equation}
\label{sigapprx}
\hat{\sigma}^+_{k'}\hat{\sigma}^-_k \approx \delta_{kk'} \hat{\sigma}^+_{k'}\hat{\sigma}^-_k,
\end{equation}
which is valid for short times when starting from the initial
state $|F\rangle$. In order to develop a classical model in
analogy with the degenerate case we define
\begin{equation}
\hat{n}^b_k = 1-\frac{1}{2}\left ( \hat{c}^\dagger_k\hat{c}_k +
\hat{c}^\dagger_{-k}\hat{c}_{-k}\right).
\end{equation}
Since $\sum_k \langle\hat{n}_k^b\rangle = \langle
\hat{n}_b\rangle$, where the sum runs over momenta inside the
Fermi sea, the expectation value of this operator can be
interpreted as that fraction of the initial pair of atoms of
momenta $(-k,k)$ that has been converted into a molecule. Note
that
\begin{equation}
\hat{\bm \sigma}_k^2=\hat{\sigma}_k^+\hat{\sigma}_k^-+\hat{\sigma}_k^z(\hat{\sigma}_k^z-1)
\end{equation}
with $\hat{\bm \sigma}_k^2|F\rangle=\frac{1}{2}(\frac{1}{2}+1)|F\rangle$
and
\begin{equation}
\label{sigz}
\hat{\sigma}_k^z = \frac{1}{2} - \hat{n}_k^b,
\end{equation}
so
\begin{eqnarray}
\nonumber
\hat{\sigma}^+_k\hat{\sigma}^-_{k}&=&\hat{\bm \sigma}_k^2-\hat{\sigma}_k^z(\hat{\sigma}_k^z-1)\\
&=& -(\hat{n}^b_k)^2 + \hat{\zeta}_k^+\hat{\zeta}_k^-, \label{sigpm}
\end{eqnarray}
where we have defined
\begin{equation}
\hat{\zeta}_k^+\hat{\zeta}_k^- = \hat{\bm \sigma}_k^2-\frac{1}{2}(\frac{1}{2}-1).
\end{equation}
Finally, by substituting Eqs. (\ref{sigpm}) and (\ref{sigz}) into Eq.
(\ref{nbk})
we obtain
\begin{equation}
\label{jyk2}
\frac{d \hat{j}_k^y}{dt}
\approx -\delta_k  \hat{j}_k^x-\chi \hat{n}_b -\chi\hat{\zeta}_k^+\hat{\zeta}_k^-.
\end{equation}
Short time results were obtained
by integrating Eqs. (\ref{nbk}), (\ref{jkx}) and (\ref{jyk2})
numerically using a fourth order Runge-Kutta procedure.
Those numerical simulations should reproduce the degenerate model
in the limit $\beta\ll1$. Note however that the short-time approximation
(\ref{sigapprx}) is slightly different from that made in order to
obtain Eq.(\ref{jy2}).  In particular, Eq. (\ref{sigapprx})
ignores a contribution linear in $\hat{n}_b$ and as a consequence
$\langle\hat{n}_b\rangle$ increases more slowly in the approximate
non-degenerate simulation so that for small $\beta$ it agrees only
within $\sim 10\%$ with the degenerate model once $5\%$ of the
initial population of atom pairs has been converted into
molecules.  This is illustrated in Fig.~\ref{fig6} which plots
the average molecule number as a function of time.

Although calculations with larger particle number are
computationally tractable for short times, we have restricted our
calculations to 18 atom pairs in order to allow an exact numerical
treatment of the full quantum non-degenerate model.
Figure~\ref{fig6} shows from left to right the degenerate
analytical solution (\ref{nbt2}), the full quantum result with
$\beta = 0.1$, the short time result with $\beta = 0.1$, short
time result with $\beta = 10$ and the full quantum result for
$\beta = 10$, respectively. The full quantum result with $\beta =
0.1$ is indistinguishable from the full quantum treatment in the
degenerate model, as expected.   These simulations illustrate in
particular that the formation of molecules is suppressed for
increased $\beta$. This is a consequence of the fact that a
significant fraction of the atom pairs are detuned from resonance
and therefore converted more slowly and incompletely into
molecules.

\section{Summary and outlook}

We have shown that the early stages of molecular dimer formation
from fermionic atoms are characterized by large fluctuations that
reflect the quantum fluctuations in the initial atomic state. In
contrast, the reverse process of dissociation of a condensate of
molecular dimers is largely deterministic. The reason for this
asymmetry can be traced to the fact that in contrast to a
quantum-degenerate fermionic system, the initial state of the
molecular condensate is well described by a mean-field theory,
that is, it is largely classical and relatively devoid of quantum
fluctuations.

As long as the atom-molecule coupling is dominant, $\beta\ll 1$,
the kinetic energies are unimportant and a degenerate model can
accurately describe the molecule formation. This is confirmed by
our numerical simulations which show that the results of the
degenerate model and the full quantum results are
indistinguishable for $\beta \lesssim 0.1$.  Small deviations
appear for $\beta \approx 1$, and the creation of molecules is
dramatically suppressed for higher ratios.

Future work will extend these considerations to a more realistic
multimode description of the association process as well as a more
detailed description of the two-body physics, which should in
particular include the dressing of molecules by the open channel
atomic pairs that is important in Feshbach sweeps. We will also
extend our model to the case of fermionic molecules.

This work is supported in part by the US Office of Naval Research,
the NSF, the US Army Research Office, NASA, and the Joint Services
Optics Program.


\begin{references}
\bibitem{Foot1} Although $^{87}$Rb is bosonic, its
statistical nature does not enter in the calculation of coupling
strength so that it gives an adequate order of magnitude estimate.
\end{references}
\end{document}